\documentstyle[fleqn,twoside]{article}

\newcommand{\AmS}{{\protect\the\textfont2
  A\kern-.1667em\lower.5ex\hbox{M}\kern-.125emS}}

\newcommand{\bel}[1]{\begin{equation}\label{#1}}
\newcommand{\ee}{\end{equation}}

\def\ramaS{\vadjust{\vbox to 0pt{\vss \hbox to \hsize
{\hskip\hsize \quad $\Leftarrow$\quad {\it SAS}\hss}\vskip3.5pt}}}
\def\ramaY{\vadjust{\vbox to 0pt{\vss \hbox to \hsize
{\hskip\hsize \quad $\Leftarrow$\quad {\it YAKR-E}\hss}\vskip3.5pt}}}
\def\rama{\vadjust{\vbox to 0pt{\vss \hbox to \hsize
{\hskip\hsize \quad $\Leftarrow$\quad
{$\Longleftarrow$}\hss}\vskip3.5pt}}}
\def\ramaM{\vadjust{\vbox to 0pt{\vss \hbox to \hsize
{\hskip\hsize \quad $\Leftarrow$\quad {\it SM}\hss}\vskip3.5pt}}}

\def\siml{\hbox{\kern.1em \lower.6ex \hbox{$\sim$} \kern-1.12em
          \raise.6ex \hbox{$<$} \kern.1em }}
\def\simg{\hbox{\kern.1em \lower.6ex \hbox{$\sim$} \kern-1.12em
          \raise.6ex \hbox{$>$} \kern.1em }}

\title{{\bf On the theory of quasielastic lepton-hadron scattering}}

\author{Ya.D. Krivenko-Emetov
\thanks{ {\it E-mail address:} yakr@kinr.kiev.ua}
\\
{\em Institute for Nuclear Research,
Ukrainian Academy of Sciences},\\ Prospect Nauki, 47,
Kiev  03680, Ukraine\\
{\em Applied Physics Department of Kiev National
Technical University},\\ Prospect Pobedu, 37,
Kiev 03056, Ukraine}

\begin{document}

\begin{abstract}
The $F_{2}$ structure functions of the inelastic lepton-hadron scattering
is calculated in the case of non-zero intermediate gluon-quarks self-energy
$M_{gq}^{2}$ and quasielastic limit. It is shown that in the quasielastic limit
the evolution equation can been solved in case massive particles as QCD ladder
calculation results.
The intermediate gluon-quarks self-energy, or polarization mass is introduced as a parameter
in two cases:1) not momentum dependence and 2) momentum dependence cases. At the
case 1) of massive particles we compare our result with Shirkov calculates (D.V.Shirkov 1999).
It has been shown (case 2.) that the proton structure function $F_{2}$
calculated at realistic values of $M_{gq}^{2}$ in high energy limit transfer
to  standard QCD ladder calculation results.

\vspace{1pc}


\end{abstract}

\maketitle

\section{INTRODUCTION} 

It is known that gluons are masless particles due to 1) principles
renormalization theory, 2) asymptotic freedom and 3) exact $SU(3)_{c}$ symmetry
and local invariance assumption. However, the question of a non-zero gluon mass  possibility
and its physical consequences is discussed for a long time in literature.
It was assumed, in particular, that gluons become massive
due to symmetry breaking \cite{1,2,Nelipa}. In this case the requirements
1) and 2) can be fulfilled. In Section 2, proposed another mechanism for the gluon
mass generation(of course our supposition dose not extenuate the significant of result \cite{1}). Gluon becomes massive due to the interaction with
quarks. In this case the quarks-gluons mass is nothing else but polarization
correction to its self-energy. This quantity is a real correction
to quarks-gluon self energy, no matter if particle is on or out
of mass surface. Therefore, this assumption does not lead to
impossibility to use Katkowski rules for the full cross section calculations
by using the optical theorem \cite{3}.
We assume that this quantity at low energy 1.)has not energy momentum
dependence and is an effective theory parameter $M^2$,
which can have anarbitrary sign, 2.) leads as $k^2L^2(k^2)$, where $L^2(k^2)$
is  $k^2$ momentum dependence and $k^2$-transfer moment of gluon
and 2.1) use parametrization \cite{garik}
put $L^2=\tilde{L}^2(ln[(\Lambda^2_N+k^2)/\mu^2_{QCD}]/ln[\Lambda^2_N/\mu^2_{QCD}])$
( $\ln(\mu^2)=ln(\Lambda^2_{QCD})+4\pi/(\beta_0\alpha_0)$),
where $\Lambda_N$- uncalculable factor that describe unperturbativ contribution in
strong coupling or quarks-gluon self-energy,
and $\Lambda_{QCD}\simeq~216(_+^-25)~ MeV$ \cite{PDG} is the constant of QCD.
In our model $M_{gq}^2$ not only gluon self-energy. This parameter also discount
(check) quarks self-energy according theory of renormalizable
(see eq. (\ref{masspl}) andless).
Its calculation can not be performed exactly since the perturbation theory
is not applicable in the QCD theory. Justification for investigate this not
momentum dependence parameter can be obtained in modern Effective Field Theory and
Quantum Hadrodynamics (see for example \cite{Wal},\cite{We90a},\cite{We91}).
This correction $M_{gq}$ must be small, but greater
with respect to quark mass. However, it can lead to some physical effects.
Namely, we show that the self-energy correction leads to a significant
change in the lepton-hadron scattering cross section.
This effect takes place because the set of ladder diagrams with infinite
number of gluon lines must be taken into account. Thus,  the intercept
correction is of order $M_{gq}^2/Q^2$, and the logarithm is of order
$M_{gq}^2(Q^2,\Lambda_{QCD}^{2})$, where $Q^2$ is the squire of transferred momentum
($Q^{2}=-q^{2}$).
Formally, the contribution
of structural function to cross section becomes important at
small $Q^{2}$, but at these moment perturabtive ladder approximation
does not work, and the polarization correction is not known so far.
It is known that at large transfer moment
polarizability contribution can be taken  into account
by renormalization of the coupling constant $g\rightarrow g(k^{2})$,
since these contributions can be factorized and by using the Ward identity
\cite{Bogol}.
In this case "polarization mass" is 
\[
M^{2}_{pol}(k^{2})\simeq_{k^2>Q_0^2}k^2\tilde{L}^2(ln[(\Lambda^2_N+k^2)/\mu^2_{QCD}]/ln[\Lambda^2_N/\mu^2_{QCD}])             2
\]
\begin{equation}
\rightarrow_{k^2\rightarrow\infty}-k^{2}b\ln(k^{2}/\mu^{2}),~(\tilde{L^2}/Ln(\Lambda^2_N/\mu^2_{QCD})\sim -b)
\label{masspl}
\end{equation}
($Im[\frac{e}{1+e\ln(k^2/\mu^2)}\frac{1}{k^2+i0}]=
e\delta(k^2+k^2e\ln(k^2/\mu^2))=e\delta(k^2-M^2)=$\\
$\frac{e}{1+e\ln(k^2/\mu^2)}\delta(k^2)$ )
where ~$b=11-(2/3)n_{f}$,$n_f$-quarks flowers \cite{Yndur}.
Comparison of the cross-section calculated in this way with experimental
results is a possible indirect way to estimate the value of possible
quarks-gluon mass \cite{1,2,4,5,6}.
It is important to say that processes of creation of other particles
 also give a contribution
to ``gluon mass'', therefore our effective $M_g$ is assumed to take
into account these contributions too.
This gives as a possibility to describe the quasiperturbative
region of scattering when $Q^{2}$ is not very large, and
maybe even when $Q^{2}$ is rather large, in contrast to the previous
results, we the polarization correction was calculated at
large $Q^{2}$ \cite{7,BalLi7877,8,9,kuraev85,10}.It is shown (eq.(\ref{transfer})), that
in case large $Q^{2}$ our results traversed to results \cite{7,BalLi7877,8,9,kuraev85,10}.
  At fig1.,2 we present the model \cite{1} (\ref{f}), and our
estimation with massive effective particles with different value of parameter and variables.
 Of course, this solution (\ref{f}),(\ref{transfer1})
is not correct at very small $Q^{2}$, when one loop approximation is incorrect.
But the one-lop solutions at large $Q^{2}$ and our solution (\ref{f}),
(\ref{transfer}) at intermediate $Q^{2}$ will be ``sewed'' at some point of transfer
momentum.(method "sewed" see for example \cite{kobkriv},\cite{garik}).
Analytically this "sewed" give formula (\ref{transfer})-(\ref{endres}).
Also, the non-singlet diagrams give the biggest contribution
at $x\rightarrow 1$, but in this case delta-functions will
give large contributions in terms which is difficult to calculate, and we did
not take into account. It was shown in \cite{16}, for example,
that these diagrams lead to a good description of experiment at $x>0.65$.
Finally, Section 3. contains a summary and conclusions.

\section{ONE-LOOP DIAGRAMS}

Let us use the axial gauge $A_{\mu}\eta^{\mu}=0$. It should be mentioned,
that in the case of massive particles the gauge is not arbitrary and the
propagator is
$$
G_{\mu\nu}=-\frac{1}{p^{2}-m^{2}+i0}(g_{\mu\nu}
-\frac{p_{\mu}p_{\nu}}{m^{2}}).
$$
As it was mentioned in the Introduction, we assume that virtual gluons
have non-zero self-energy, or polarization mass $M_{g}(Q^{2})$.
In this case the hadron tensor has the next form \cite{ChengLi}
($S_2(V)$-color multiply, M-mass of hadron)
\begin{eqnarray}
W_{\mu\nu}(p,q)=\frac{g^{2}S_2(V)}{2M}
\int \frac{d^{4}k}{(2\pi)^{3}} \times \nonumber \\
\frac{\delta [(p-k)^{2}+M_{gq}^{2}]\delta [(q+k)^{2}]}{k^{4}}
d^{\rho\sigma}(p-k)T_{\rho\mu\nu\sigma}+\Delta W_{\mu\nu}.
\label{Tensor}
\end{eqnarray}
Where $\Delta W_{\mu\nu}$ discount, that $M^2_{gq}$ can have more
composite structure, then constant or $k^2L^2$.
Here we substitute the quark and gluon propagator by delta-functions
with arguments equal to the denominator of propagator expressions.

Let us introduce the Sudakov variables:
$$
k_{\mu}=\xi p_{\mu}+\beta q'_{\mu}+k_{\perp\mu}
$$
with
$$
q'^{2}\simeq p^{2}\simeq 0, \ \ \ \ pq'\simeq pq,
$$                                              
and a new variable $x$: $q'=q+xp$.
In this case the integral in (\ref{Tensor}) can be transformed as
$$
\int d^{4}k\delta [(p-k)^{2}-M_g^{2}]\delta [(q+k)^{2}]=
$$
$$
=2\pi pq\int d\beta d\xi d k_{\perp}^{2}
\frac{1}{2pq(1-\xi )}
\delta [\beta -\frac{k_{\perp}^{2}-M_{gq}^{2}}{2\pi pq(1-\xi )}]
$$
\begin{equation}
\times\frac{1}{2pq}
\delta[\xi-x+\frac{(1-x)k^2}{2pq}+\frac{M^2x}{2pq}]
\label{delta}
\end{equation}
(see Appendix) where we use the substitution
$d^{4}k =2\pi pqd\beta d\xi d k_{\perp}^{2}$ and approximation
$-\frac{k_{\perp}}{q'}+\beta^2\frac{q'^2}{2pq(1-\xi)}\rightarrow~0$.\\
If
\begin{equation}
\beta<1-\xi~~=>\xi\sim~x
\label{approximation}
\end{equation}
,where
\begin{equation}
\xi\simeq x_{f}\simeq\frac{2x_{b}}{1+\sqrt{1+4M^{2}x_{b}^{2}/Q^{2}}}
\label{xfeim}
\end{equation}
(\cite{berken},\cite{16},
when $Q^{2}>>M^{2}_{n}$,~~~$\xi\rightarrow x_{b}=x$).
Similarly to \cite{7,9}, the appearance of the term $ln(-q^2/p^2)$
is related to the phase space regions, where the relations
(\ref{approximation}) are correct.
Therefore, variable $x$ which corresponds to the parton
model variable ia part of longitudinal momentum and it justifies
the approximation $q'^{2}\simeq 0$.

The square of $k_{\mu}$ can be expressed as
$
k^2=\xi^2 p^2+\beta\xi p^{\mu}q'_{\mu}+\xi p^{\mu}k_{\perp,\mu}+
\beta\xi q'^{\mu}p_{\mu}+\beta^2 q'^2+\beta q'^{\mu}k_{\perp,\mu}+
\xi k^{\mu}_{\perp}p_{\mu}+
\beta k_{\perp,\mu}q'^{\mu}+k_{\perp}^2\simeq~2\beta\xi(q'p)+k_{\perp}^2
$ use (\ref{delta}) , obtain
\begin{equation}
k^{2}\simeq \frac{k_{\perp}^{2}}{(1-\xi )}-\frac{M_{gq}^{2}\xi}{(1-\xi )}.
\label{k2}
\end{equation}

Convolution of the hadron and gluon polarization
tensors with putting $\eta =q'$ and using of $\gamma$-matrix algebra,
(\ref{delta}),(\ref{k2})
gives the main contribution to the hadron tensor, which corresponds
to the tree diagram. If $\beta<1-\xi~~=>\xi\sim~x$ , obtain
\begin{eqnarray}
d^{\rho\sigma}(p-k)T_{\rho\mu\nu\sigma}-\Delta W_{\mu\nu}=
2tr[\gamma_{\mu}({\hat q}+{\hat k})\gamma_{\nu}]
(-{\hat k}{\hat p}{\hat k}+\nonumber \\
+\frac{k^{2}}{(p-k,q')}
[(pq'){\hat k}+(kq'){\hat p}+(p-k,p){\hat q'}])
\label{dTlast}
\end{eqnarray}
and use (\ref{M2q})-(\ref{trbeta<xi})(see Appendix), obtain at case\\
1.1) $M^2(k^2)\simeq~M^2$
\begin{eqnarray}
\simeq~2k^2tr[{\hat p}\gamma_{\mu}({\hat q}+{\hat k})\gamma_{\nu}]
\left(\frac{1+\xi^{2}}{1-\xi}+\frac{M_{g}^{2}\xi}{pq}
+\frac{M_{g}^{2}\xi}{pq(1-\xi)}
-\frac{M_{g}^{4}\xi}{2pqk^{2}}\right)
\label{dT}
\end{eqnarray}
2.1) and at case  $M^2(k^2)\simeq~L^2k^2$ (supply it at (\ref{trbeta<xi}),obtain)
\begin{eqnarray}
\simeq~2k^2tr[{\hat p}\gamma_{\mu}({\hat q}+{\hat k})\gamma_{\nu}]
\left(\frac{1+\xi^{2}}{1-\xi}+L^2\xi\right)
\label{dT7}
\end{eqnarray}
and if $\beta>1-\xi~~=>\xi\sim~x-\frac{(1-x)k^2}{2pq}-\frac{M^2x}{2pq}$
(see Appendix, eq.(\ref{trbeta>xi}))
\[
d^{\rho\sigma}(p-k)T_{\rho\mu\nu\sigma}-\Delta W_{\mu\nu}\simeq~2
tr[{\hat p}\gamma_{\mu}({\hat q}+{\hat k})\gamma_{\nu}]
\]
\begin{equation}
\left
(-\frac{(k^{2}-M_{g}^{2})^{2}\xi}{2pq}
+M_{g}^{2}\xi+k^2(1-x)+
\frac{k^{2}2\xi}{\beta}-
\frac{k^{2}2\xi(1-\xi)}{\beta}
\right)
\label{labda3}
\end{equation}
2.2.) if $M^2=k^2L^2~~~~~,\beta>1-\xi~~=>\xi\sim~x-
\frac{(1-x)k^2}{2pq}-\frac{k^2L^2x}{2pq}$ (see Appendix, eq.
(\ref{beta>xi1}))
\begin{equation}
(eq.\ref{labda3})\simeq 2k^2
tr[{\hat p}\gamma_{\mu}({\hat q}+{\hat k})\gamma_{\nu}]\left
(L^{2}x+(1-x)-\frac{2(x-x^2+L^2x^2)}{(1-L^2)}
\right)
\label{labda4}
\end{equation}
 To obtain the expression (\ref{dT}),(\ref{labda3}),(\ref{labda4}) the
first and the second terms (equal to $(1-\xi )k^{2}$ and $2\xi k^{2}/(1-\xi )$,
correspondingly) in gluon propagator were considered.
 We also neglecting $1/k^{4}$ and $k^4$, because
at high $k^2$~~ $|M_{g}^{2}|\simeq~k^{2}ln(k^{2}/\Lambda^2_{QCD})>k^2$, and
$k^2/2pq=k^2x/Q^2\simeq k^2L^2/Q^2_{Q^2\rightarrow\infty}\rightarrow~0$ .
 However, in contrast to the paper \cite{7,9},
first and the third term in (\ref{hatkpk}) was also taken
into account, since it gives the contribution
from the polarization mass in (\ref{dT}) and (\ref{dT7}) respectively.
  If~~~$\beta<1-\xi~~=>\xi\sim~x$ use (\ref{delta}),(\ref{k2}),(\ref{dT}),
  obtain in cases\\
1.1.) $M^2(k^2)\simeq M^2$
$$
(\ref{intdelt})=\frac{3g^{2}S_2(V)}{2M_n(2\pi)^{3}}
\pi\int_{m^2-f(M^2)}^{Q^2-f(M^2)}\frac{d\beta d\xi dk^{2}}{k^{2}}
~2tr[{\hat p}\gamma_{\mu}({\hat q}+{\hat k})\gamma_{\nu}]\times
$$
$$
\times\left(\frac{1+\xi^{2}}{1-\xi}+\frac{M_{g}^{2}\xi}{pq}
+\frac{M_{g}^{2}\xi}{pq(1-\xi)}
-\frac{M_{g}^{4}\xi}{2pqk^{2}}\right)\times
$$
\begin{equation}
\times\frac{1}{2pq}\delta [\beta-
(k^{2}-M_{g}^{2})/2pq]
\delta((\xi-x)+\frac{(1-x)k^2}{2pq}+\frac{M^2(x-\xi^2)}{2pq})
\label{finalm2}
\end{equation}
2.1.)According (\ref{masspl}), put $M^2\simeq k^2L^2$
\[
=\frac{3g^{2}S_2(V)}{2M_n(2\pi)^{3}2pq}
\pi\ln
\left[\frac{Q^2-Q^2L^2x/(1-x)}{m^2-m^2L^2x/(1-x)}\right]
~2[8\xi p_{\nu}p_{\mu}-
\]
\begin{equation}
4M\nu g_{\mu\nu}+...]
\left(\frac{1+x^{2}}{1-x}+A^2x+\frac{A^2x}{1-x}(\frac{1+x^{2}}{1-x}+A^2x)\right)
\label{kl2}
\end{equation}
According (\ref{finalm2}) we obtain with
equations (\ref{Tensor}) and (\ref{dT}) give (if~$\beta<1-\xi,~~~
M^2(k^2)=M^2$)
\[
f(x,Q^{2})-\Delta f(x,Q^2) \sim g^2\frac{M_{g}^{4}x^{2}}{Q^{4}}+
\]
\begin{equation}
+ g^{2}ln[\frac{Q^{2}-M_{g}^{2}~x/(1-x)}{m^{2}
+-M_{g}^{2}~x/(1-x)}]
(\frac{1+x^{2}}{1-x}
+\frac{2M_{g}^{2}x^{2}}{Q^{2}}
+\frac{2M_{g}^{2}x^{2}}{Q^{2}(1-x)})
\label{f}
\end{equation}
(for massive gluons $\Delta f(x,Q^2)=0$ and 
we have put $x=Q^{2}/2pq$ in the last expression),
and if $M^2=k^2L^2,~~~\beta<1-\xi$ (\ref{kl2}) give (put \\
$L^2=\tilde{L}^2(ln[(\Lambda^2_N+Q^2)/\Lambda^2_{QCD}]/ln[\Lambda^2_N/\Lambda^2_{QCD}])$,~$|ln(\Lambda^2_{QCD})|>>4\pi/(\beta_0\alpha_0)$ )
\[
f(x,Q^{2})-\Delta f(x,Q^2)
\sim g^2\ln\left[\frac{Q^2-Q^2L^2x/(1-x)}{m^2-m^2L^2x/(1-x)}\right]
\]
\begin{equation}
\left(\frac{1+x^{2}}{1-x}+A^2x+\frac{A^4x^2}{(1-x)}
+\frac{A^2x(1+x^{2})}{(1-x)^2}\right)\rightarrow_{Q^2\rightarrow\infty}
\label{transfer1}
\end{equation}
\[
\rightarrow_{Q^2\rightarrow\infty}
g^2\left(\ln(Q^2/m^2)+ln[\frac{Ln(Q^2/\Lambda^2_{QCD})}{Ln(m^2/\Lambda^2_{QCD})}]\right)\times
(\frac{1+x^{2}}{1-x}+
\]
\begin{equation}
\tilde{A}^2\left(\frac{
x(1+x^{2})}{(1-x)^2}+x+\frac{\tilde{A^2}x^2}{(1-x)}\right))
\label{transfer}
\end{equation}
 Compare  this result with standard result at high $Q^2$, obtain  that
 at $Q^2\rightarrow\infty$
\[
\Delta~f(x,Q^2)\simeq
-g^2\ln(Q^2/m^2)[
\left(\frac{1+x^{2}}{1-x}\right)+1+
\frac{\tilde{A^2}}{g^2\ln(Q^2/m^2)}
\]
\begin{equation}
\times Ln[\frac{Ln(Q^2/\Lambda^2_{QCD})}{Ln(m^2/\Lambda^2_{QCD})}]]
\left(\frac{
x(1+x^{2})}{(1-x)^2}+x+\frac{\tilde{A^2}
x^2}{(1-x)}\right)
\label{stres}
\end{equation}
and at $L^2\rightarrow 0$~~$\Delta~f(x,Q^2)\simeq 0$
But this result(eq. (\ref{transfer})) give badly behavior at $L\rightarrow~0$($A^2\rightarrow~0$),
because at $Q^2+\Lambda^2_N=\Lambda^2_{QCD}$ must be fulfillment condition
$m^2=\Lambda^2_{QCD}-\Lambda_N^2$. Good behavior
have another renormalizable result, that received similarly
(we know polarized mass only at high $k^2$~.This unknown function we can transform
into Teilor series $M^2=const^{+}_{-}(Q^2/Q^2_{unpert})f(m^2)^{-}_{+}...\rightarrow_{Q^2>>Q^2_{unpert}}
Q^2\ln(m^2/\Lambda^2_{QCD}),Q^2_{unpert}\sim 3-5 Gev^2 $~
,$L^2$ gives (\ref{masspl})):
\[
f(x,Q^2)\simeq~const*\frac{48g^{2}S_2(V)}{2(2\pi)^{3}}
\pi[\ln\left[\frac{Q^2-Q^2L^2(Q^2)x/(1-x)}{m^2-Q^2\tilde{f(m^2)}
x/(1-x)}\right]\times
\]
\[
\times\left(\frac{1+x^{2}}{1-x}+A^2x+\frac{A^2x}{1-x}
(\frac{1+x^{2}}{1-x}+A^2x)\right)
-Ln[\frac{Ln(Q^2/\Lambda^2_{QCD})}
{Ln(m^2/\Lambda^2_{QCD})}]\times
\]
\begin{equation}
\times\tilde{A}^2\left(x+\frac{x}{1-x}
(\frac{1+x^{2}}{1-x}+\tilde{A}^2x)\right)
\label{endres}
\end{equation}

\section{CONCLUSIONS}

It has been shown that the contribution to  inelastic lepton-hadron
scattering cross section from the quarks-gluon polarization mass can
be rather large.
Polarization mass consideration can be a possible
way to decrease the difference between experimental data and
theoretical calculations of the structure functions  of  inelastic
hadron lepton scattering. Another application of the
results obtained in this paper is a possible estimation of
the quarks-gluon self-energy  results calculated by formula
(\ref{f}) (\ref{endres}), which contains quarks-gluon self-energy
as a function-parameter. We believe that quarks-gluon
polarization mass presence will make the condition
$k_1<<k_2<<...<<k_N$ weaker \cite{yakre},
even if these other processes are not taken into account.
And that this gives as a possibility to describe the quasiperturbative
region of scattering when $Q^{2}$ is not very large, and
maybe even when $Q^{2}$ is rather large, in contrast to the previous
results, where the polarization correction was calculated at
large $Q^{2}$ \cite{7,BalLi7877,8,9,kuraev85,10}.
  At fig.1.,2. we present our calculation for the model \cite{1} (\ref{f}),
and our estimation (\ref{endres}) with effective polarization mass with
different value of parameter and variables. How apparently at fig.1 "skailing"
increase with increase mass of gluons. In fig.2 "skailing" is more visible and
in principle can be used for describe experimental data.

\section*{Acknowledgments}

Ya.D.K.-E. would like to thank  V.M. Turkowski, A.P.Kobushkin
and V.V.Davidovsky for useful discussions
which have resulted in an improvement of the paper.
Ya.D.K.-E. acknowledges the financial support from the
Applied Physics Department of Kiev National Technical University,
Project N2700/162.He also thanks V.M.Kolomiec
for aid.

\section*{APPENDIX}
For obtain (\ref{delta}), use \\
$\delta ((p-k)^{2}-M_g^{2})=\delta (\beta^2q'^2+k_{\perp}^2-2pq'\beta(1-\xi)
-2pk_{\perp}(1-\xi)-M_{gk}^2)$
$
\simeq~\frac{1}{2pq(1-\xi)}\delta (-\beta+\frac{k_{\perp}^{2}-M_{g}^{2}}
{2\pi pq(1-\xi)}-\frac{k_{\perp}}{q'}+\beta^2\frac{q'^2}{2pq(1-\xi)})
\simeq~\frac{1}{2pq(1-\xi)}\delta (-\beta+\frac{k_{\perp}^{2}-M_{g}^{2}}
{2\pi pq(1-\xi )})$, and\\
$\delta (k+q)^2\simeq~\delta((\beta+1)q'+(\xi-x)p+k_{\perp})^2\simeq~
\delta(2(\beta+1)(\xi-x)pq+k_{\perp}^2)
\simeq\frac{1}{2pq}\delta
((\xi-x)-\beta~x+\frac{k^2}{2pq}-\frac{k^2\xi}{2pq}+\beta\xi
+\frac{M^2\xi(1-\xi)}{2pq})
\simeq\frac{1}{2pq}\delta
((\xi-x)+\frac{(1-x)k^2}{2pq}+\frac{M^2(x-\xi^2)}{2pq})$
\begin{equation}
\simeq~\frac{1}{2pq}\delta(\xi-x+\frac{(1-x)k^2}{2pq}+\frac{M^2x}{2pq})
=\frac{1}{2pq}\delta(\xi-x+\frac{(1-x)k^2x}{Q^2}+\frac{M^2x^2}{Q^2})
\label{delt2}
\end{equation}
Supplied 
\begin{equation}
-(p-(p-k))^{2}=-p^{2}-(p-k)^{2}+2p(p-k)=-p^{2}-M_{g}^{2}+2p(p-k)
\label{M2q}
\end{equation}
and
$(p-k)^{2}=M_{g}^{2}\rightarrow ~
p(p-k)=p^{2}/2-k^{2}/2+M^{2}_{g}/2
\simeq~p^{2}/2+M_{g}^{2}/(2(1-\xi))-
k_{\perp}^{2}/(1-\xi)
$
in to (\ref{dTlast}) , obtain
\begin{eqnarray}
d^{\rho\sigma}(p-k)T_{\rho\mu\nu\sigma}=2tr[\gamma_{\mu}({\hat q}
+{\hat k})\gamma_{\nu}]~~~~~~~~~~~\nonumber\\
\times~(-{\hat k}{\hat p}{\hat k}+\frac{k^{2}}{(p-k,q')}
[(pq'){\hat k}+(kq'){\hat p}+(p^{2}/2+M_{g}^{2}/2
-k^{2}/2){\hat q'}])
\label{trase}
\end{eqnarray}
and use
\begin{equation}
   {\hat k}{\hat p}{\hat k}=
   2\beta(pq)(\beta~{\hat q'}+k_{\perp})-
   (k^{2}(1-\xi)+M_{g}^{2}\xi){\hat p}
\label{hatkpk}
\end{equation}
$(p-k)q'=(1-\xi+\beta)\frac{k^{2}-k_{\perp}^{2}}{2\beta~\xi}$,~~
$(pq')\hat k+(kq')\hat p=\frac{k^{2}-k_{\perp}^{2}}{2\beta~\xi}(\hat k+\xi\hat p)$ ,\\
$ k^{2}=2\beta\xi(pq)+k_{\perp}^{2} $,~
$\beta\simeq~(k_{\perp}^{2}-M_{g}^{2})/(2pq(1-\xi))=(k^{2}-M_{g}^{2})/2pq$
rewrite (\ref{trase}) as
\[
\simeq~2tr[...]
(-2\beta(pq)\beta{\hat q'}+(k^{2}(1-\xi)+M_{g}^{2}\xi)\hat p+
\frac{k^{2}}{1-\xi+\beta}(\hat k+\xi\hat p-
\frac{k^{2}-M_{g}^{2}}{2pq}\hat q'))
\]
\[
\simeq~2tr[...]
\left(-2\beta(pq)\beta\hat q'+(k^{2}(1-\xi)+M_{g}^{2}\xi)\hat p+
\frac{k^{2}}{1-\xi}(\beta\hat q'+2\xi\hat p-\beta\hat q')-
\frac{k^{2}2\beta\xi\hat p}{1-\xi}\right)
\]
\begin{equation}
\simeq~2tr[...]
\left(k^{2}\frac{1+\xi^{2}}{1-\xi}-\frac{(k^{2}-M_{g}^{2})^{2}\xi}{2pq}
-\frac{(k^{2}-M_{g}^{2})k^{2}\xi}{pq(1-\xi)}+M^{2}_{g}\xi\right)\simeq~
(eq.~\ref{dT})
\label{trbeta<xi}
\end{equation}
if $\beta<1-\xi$ , and if  $\beta>1-\xi$, obtain 
\[
\simeq~2tr[...]
(-2\beta(pq)\beta\hat q'+(k^{2}(1-\xi)+M_{g}^{2}\xi)\hat p+
\frac{k^{2}}{1-\xi+\beta}2\xi\hat p)
\]
\begin{equation}
\simeq~2
tr[{\hat p}\gamma_{\mu}({\hat q}+{\hat k})\gamma_{\nu}]\left
(-\frac{(k^{2}-M_{g}^{2})^{2}\xi}{2pq}
+M_{g}^{2}\xi+k^2(1-\xi)+
\frac{k^{2}2\xi}{1-\xi+\beta}\right)=(eq. (\ref{labda3}))
\label{trbeta>xi}
\end{equation}
1.) if $M^2=k^2L^2~~~~~,\beta>1-\xi$
\begin{equation}
\simeq~2
tr[{\hat p}\gamma_{\mu}({\hat q}+{\hat k})\gamma_{\nu}]\left
(k^2L^{2}x+k^2(1-x)+
\frac{2\xi Q^2}{(1-L^2)x}-
\frac{2\xi (1-\xi)Q^2}{(1-L^2)x}
\right)=(eq.(\ref{labda4}))
\label{beta>xi1}
\end{equation}
At the case $M^2(k^2)\sim~M^2$,and $\beta<1-\xi$ supply
(\ref{delta}),(\ref{dT}) and (\ref{k2}) into (\ref{Tensor}), obtain
$$
\frac{3g^{3}}{2M_n
(2\pi)^{3}}
2\pi pq\int\frac{d\beta d\xi d k_{\perp}^{2}}{k^{4}}
~2k^2tr[{\hat p}\gamma_{\mu}({\hat q}+{\hat k})\gamma_{\nu}]
\left(\frac{1+\xi^{2}}{1-\xi}+\frac{M_{g}^{2}\xi}{pq}
+\frac{M_{g}^{2}\xi}{pq(1-\xi)}
-\frac{M_{g}^{4}\xi}{2pqk^{2}}\right)
$$
\[
\frac{1}{2pq(1-\xi)}\delta [\beta-
\frac{k_{\perp}^{2}-M_g^2}{2\pi~pq(1-\xi )}]
\times\frac{1}{2pq}
\delta ((\xi-x)+\frac{(1-x)k^2}{2pq}+\frac{M^2(x-\xi^2)}{2pq})
\]
$$
=\frac{3g^{3}}{2M_n(2\pi)^{3}}
\pi\int_{m^2-f(M^2)}^{Q^2-f(M^2)}\frac{d\beta d\xi d(k^{2}+\frac{M^2\xi}{1-\xi})}{k^{2}}
~2tr[{\hat p}\gamma_{\mu}({\hat q}+{\hat k})\gamma_{\nu}]
\left(\frac{1+\xi^{2}}{1-\xi}+\frac{M_{g}^{2}\xi}{pq}
+\frac{M_{g}^{2}\xi}{pq(1-\xi)}
-\frac{M_{g}^{4}\xi}{2pqk^{2}}\right)
$$
\begin{equation}
\frac{1}{2pq}\delta [\beta-
(k^{2}-M_{g}^{2})/2pq]
\delta((\xi-x)+\frac{(1-x)k^2}{2pq}+\frac{M^2(x-\xi^2)}{2pq})
\label{intdelt}
\end{equation}

2.)According (\ref{masspl}), put $M^2\simeq k^2L^2$ into (\ref{Tensor}) and use
(\ref{dT7}),obtain
$$
=\frac{3g^{3}}{2M_n(2\pi)^{3}}
\pi\int_{m^2-\tilde{f(M^2,Q^2)}}^{Q^2-f(M^2,Q^2)}\frac{d\beta d\xi d( k^{2}+k^2\frac{L^2\xi}{1-\xi})}{k^{2}}
~2tr[{\hat p}\gamma_{\mu}({\hat q}+{\hat k})\gamma_{\nu}]
\left(\frac{1+\xi^{2}}{1-\xi}+L^2\xi\right)
$$
\[
\frac{1}{2pq}\delta [\beta-
(k^{2}-M_{g}^{2})/2pq]
\delta((\xi-x)+\frac{(1-x)k^2}{2pq}+\frac{M^2(x-\xi^2)}{2pq})
\]
$$
=\frac{3g^{3}}{2M_n(2\pi)^{3}}
\pi\int_{m^2-m^2L^2(m^2)\xi/(1-\xi)}^{Q^2-Q^2L^2(Q^2)\xi/(1-\xi)}\frac{d\beta d\xi dk^2(1+\frac{L^2\xi}{1-\xi})}{k^{2}}
~2tr[{\hat p}\gamma_{\mu}({\hat q}+{\hat k})\gamma_{\nu}]
\left(\frac{1+\xi^{2}}{1-\xi}+L^2\xi\right)
$$
\[
\frac{1}{2pq}\delta [\beta-
(k^{2}-M_{g}^{2})/2pq]
\delta((\xi-x)+\frac{(1-x)k^2}{2pq}+\frac{M^2(x-\xi^2)}{2pq})
\]
$tr[{\hat p}\gamma_{\mu}({\hat q}+{\xi\hat p})\gamma_{\nu}]=4[p_{\mu}(\xi p+q)_{\nu}
+(\xi p+q)_{\mu}p_{\nu}-p(\xi p+q)g_{\mu\nu}]=
[8\xi p_{\nu}p_{\mu}-4M\nu g_{\mu\nu}+...]$
\[
\sim\frac{24g^{2}S_2(V)\pi}{M_n(2\pi)^{3}}
\ln\left[\frac{Q^2-Q^2L^2x/(1-x)}{m^2-m^2L^2x/(1-x)}\right](1+\frac{A^2x}{(1-x)})
\]
\begin{equation}
[\xi\frac{p_{\nu}p_{\mu}}{8*2pq}-\frac{2M\nu g_{\mu\nu}}{4*2pq}+..]
\left(\frac{1+x^{2}}{1-x}+A^2x\right)=(\ref{kl2})
\label{fin}
\end{equation}
where  $A^2\sim -b\ln\left[\frac{\Lambda^2_N+Q^2-Q^2L^2x/(1-x)}{\Lambda^2_n+m^2-m^2L^2x/(1-x)}
\right]\rightarrow_{Q^2\rightarrow\infty}-b Ln[\frac{Ln(Q^2/\Lambda^2_{QCD})}
{Ln(m^2/\Lambda^2_{QCD})}]=\tilde{A^2}$

\section*{Figure caption}
Fig.1 Our calculation for structural function $F_{2}$ in Shirkov approximation
\cite{1} (\ref{f}) is presented as a function
of $Q^{2}$ at different values of sign of the effective massive
particles square $M_{p}^{2}$ 
and in the case of massless particles (\ref{transfer})-(\ref{stres}).\\

Fig.2 Structural function $F_{2}$ in our approximation
(\ref{endres}) is presented as a function
of $Q^{2}$ at different values of parameters of the
effective "polarized masse"  $M_{g}^{2}$($\Lambda^2_N=Labda=3 GeV^2$).
 This results compare with well known perturbative result
(\ref{transfer})-(\ref{stres}) for massless gluons.
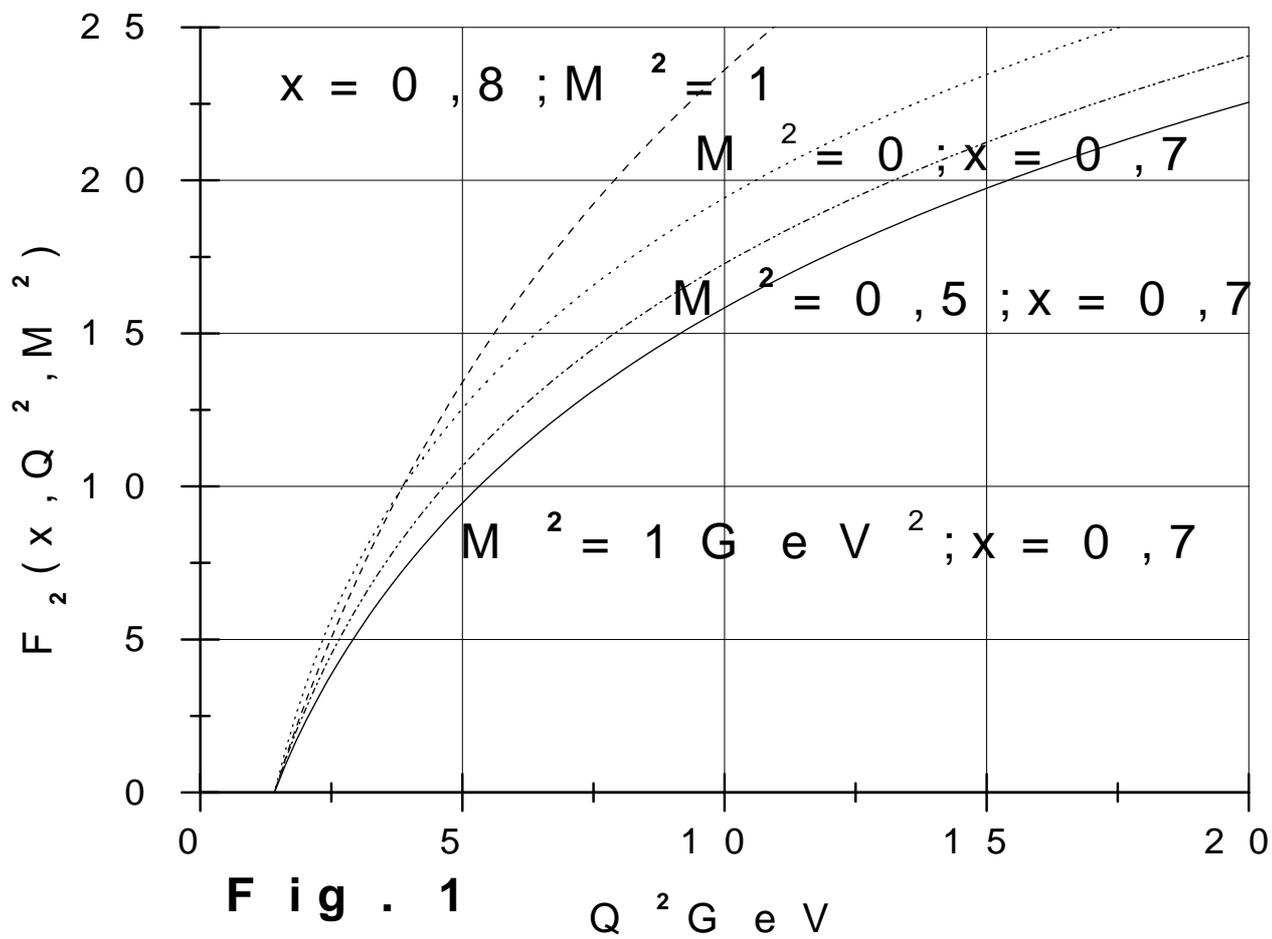
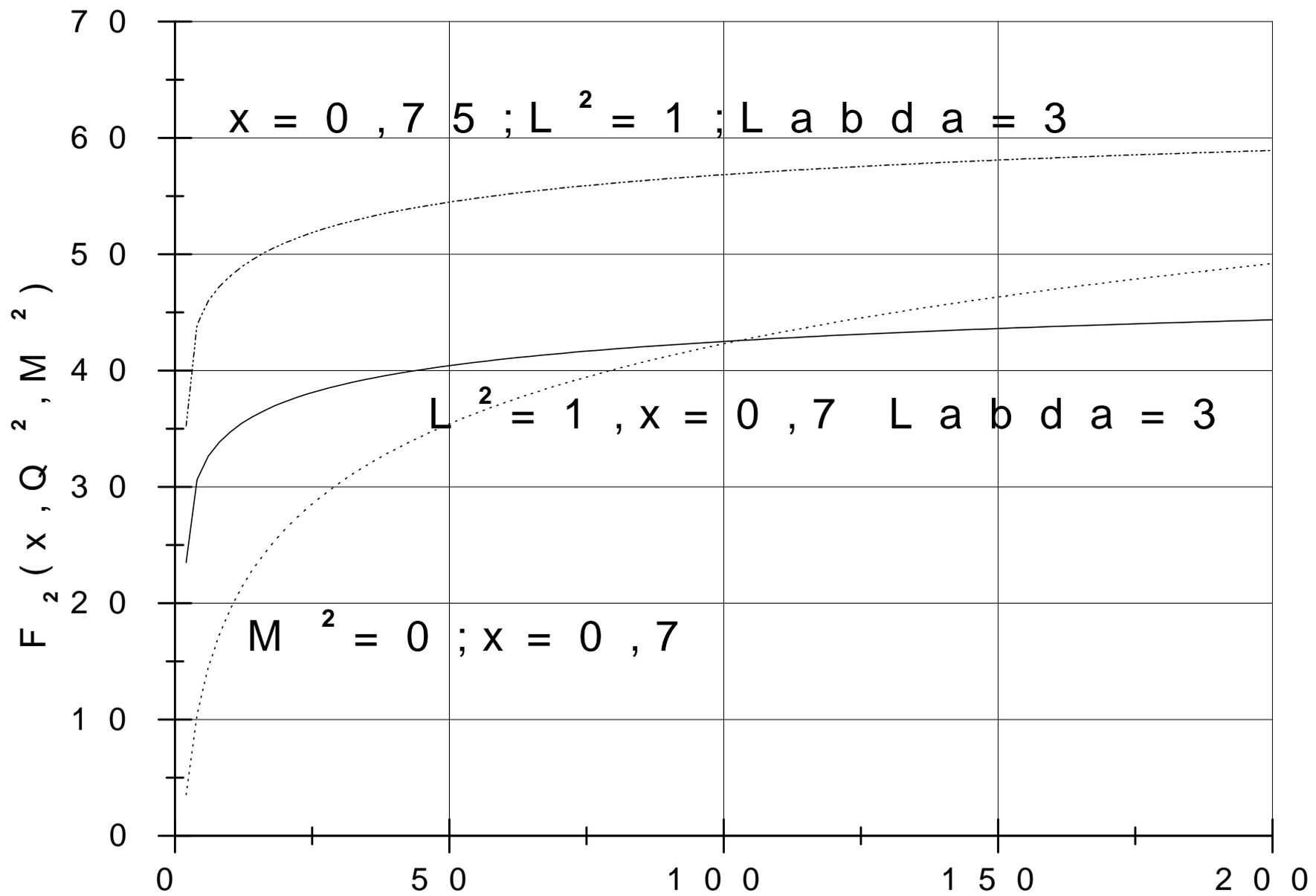
\end{document}